\documentclass[natbib]{emulateapj}

\usepackage[]{natbib}
\newcommand{\vrad}{$v_R$}

\newcommand{\logl}{log($L/{L_{\odot}}$)}

\newcommand{\loglxlbol}{log($L_{\rm X}$/$L_{bol}$)}
\newcommand{\teff}{$T_{\rm eff}$}

\newcommand{\pmra}{$\mu_{\alpha *}$}
\newcommand{\pmdec}{$\mu_{\delta}$}
\newcommand{\kms}{km\,s$^{-1}$}
\newcommand{\masyr}{mas\,yr$^{-1}$}

\newcommand{\mjup}{$M_{Jup}$}
\newcommand{\msun}{$M_{\odot}$}
\newcommand{\rsun}{$R_{\odot}$}
\newcommand{\lsun}{$L_{\odot}$}

\newcommand{\wmsq}{W\,m$^{-2}$}
\newcommand{\logg}{log($g$)}
\newcommand{\ergs}{erg\,s$^{-1}$}

\slugcomment{Astrophysical Journal Letters: Received 2012 June 7; accepted 2012 June 27}
\shorttitle{The Fomalhaut-TW PsA System}
\shortauthors{Mamajek}
\begin{document}

\title{On the Age and Binarity of Fomalhaut}

\author{Eric E. Mamajek\altaffilmark{1}} 
\affil{Cerro Tololo Inter-American Observatory,
Casilla 603, La Serena, Chile}
\altaffiltext{1}{On leave, University of Rochester, Department of Physics \& 
Astronomy, Rochester, NY, 14627-0171, USA} 
\email{emamajek@ctio.noao.edu}

\begin{abstract} 
The nearby (d = 7.7 pc) A3V star Fomalhaut is orbited by a resolved dusty debris disk and a controversial candidate extrasolar planet. The commonly cited age for the system (200\,$\pm$\,100 Myr) from \citet{Barrado97} relied on a combination of isochronal age plus youth indicators for the K4V common proper motion system TW PsA.  TW PsA is 1$^{\circ}$.96 away from Fomalhaut, and was first proposed as a companion by Luyten (1938), but the physicality of the binarity is worth testing with modern data.  I demonstrate that TW PsA is unequivocally a physical stellar companion to Fomalhaut, with true separation 0.280$^{+0.019}_{-0.012}$ pc (57.4$^{+3.9}_{-2.5}$ kAU) and sharing velocities within 0.1\,$\pm$\,0.5 km s$^{-1}$ -- consistent with being a bound system.  Hence, TW PsA should be considered ``Fomalhaut B''. Combining modern HR diagram constraints with four sets of evolutionary tracks, and assuming the star was born with protosolar composition, I estimate a new isochronal age for Fomalhaut of 450\,$\pm$\,40 Myr and mass of 1.92\,$\pm$\,0.02 M$_{\odot}$. Various stellar youth diagnostics are re-examined for TW PsA. The star's rotation, X-ray emission, and Li abundances are consistent with approximate ages of 410, 380, and 360 Myr, respectively, yielding a weighted mean age of 400\,$\pm$\,70 Myr. Combining the independent ages, I estimate a mean age for the Fomalhaut-TW PsA binary of 440 $\pm$ 40 Myr.  The older age implies that substellar companions of a given mass are approximately one magnitude fainter at IR wavelengths than previously assumed.
\end{abstract}


\section{Introduction}

Fomalhaut ($\alpha$ PsA, HD 216956, HIP 113368) is a famous nearby A3V
star with a large resolved dusty debris disk \citep[e.g.][]{Holland03}
and an imaged candidate extrasolar planet \citep{Kalas08}. The age of
Fomalhaut is mainly of interest for predicting the infrared
brightnesses of substellar companions \citep{Kenworthy09, Janson12},
calculations of the total mass of the parent bodies generating the
dust \citep{Chiang09}, and placing the dusty debris disk in
evolutionary context with other stars \citep[e.g.][]{Rieke05}.  In
general, accurate ages for host stars of substellar objects are useful
for constraining not only the masses of the companions, but accurate
ages for the youngest stars may help constrain the initial conditions
for the substellar objects \citep{Spiegel12}.  \citet{Kalas08}
recently announced the discovery of a faint optical companion (likely
$\lesssim$3 \mjup\,) at separation 12''.7 (96 AU) from Fomalhaut.
While the companion has been imaged multiple times at optical
wavelengths, \citep[][Kalas et al., in prep]{Kalas08}, it has eluded
detection in the infrared \citep{Janson12}.

Given the importance of Fomalhaut as a benchmark resolved debris disk
system and possible planetary system, a detailed reassessment of its
age is long overdue.  This paper is split into the following sections:
1) a review of published age estimates for Fomalhaut, 2) estimation of
a modern isochronal age for Fomalhaut, 3) demonstration of the
physicality of the Fomalhaut-TW PsA binary system, 4) age estimates
for TW PsA based on multiple calibrations, and 5) estimation of a
consensus age for the Fomalhaut-TW PsA system. These results supersede
the age analysis for the Fomalhaut-TW PsA system presented at the 2010
Spirit of Lyot meeting in Paris \citep{Mamajek10}.

\begin{deluxetable}{ccccc}
\tabletypesize{\scriptsize}
\tablecaption{Previous Ages for Fomalhaut\label{tab:ages}}
\tablewidth{0pt}
\tablehead{
\colhead{Age (Myr)} & \colhead{Ref} & \colhead{Method}}
\startdata
200$\pm$100	    & 1 & isochrones (Fom), Li, X-ray, rotation (TW)\\
224$^{+115}_{-119}$ & 2 & isochrones (Fom)\\
156$^{+188}_{-106}$ & 3 & isochrones (Fom)\\
290		    & 4 & isochrones (Fom)\\
480		    & 5 & isochrones (Fom)\\
220                 & 6 & ``Fomalhaut''\\
419$\pm$31          & 7 & isochrones (Fom)
\enddata
\tablecomments{``TW'' = TW PsA, ``Fom'' = Fomalhaut. References:
(1) \citet{Barrado97}, \citet{Barrado98}, 
(2) \citet{Lachaume99},
(3) \citet{Song01},
(4) \citet{diFolco04},
(5) \citet{Rieke05},
(6) \citet{Rhee07}, 
(7) \citet{Zorec12}.}
\end{deluxetable}

\section{Review of Previous Age Estimates}

Previously published ages for Fomalhaut and TW PsA are listed in Table
\ref{tab:ages}, and span a factor of 3, from 156 Myr \citep{Song01} to
480 Myr \citep{Rieke05}.  The most often cited age for Fomalhaut is
200\, $\pm$\, 100 Myr from \citet{Barrado97} and \citet{Barrado98}.
The \citet{Barrado97} estimate comes from multiple age indicators
(including Li abundance, rotation, HR diagram position, and X-ray
emission) for its purported common proper motion companion TW PsA.
Later, \citet{Barrado98} assigned the same age to Fomalhaut based on
its purported membership to the Castor Moving Group (CMG; which
included Castor, Vega, and roughly a dozen other systems).  These
analyses relied heavily on a few assumptions, worth reexamining --
namely that Fomalhaut and TW PsA are physically related, that the
Castor group is physical (i.e. useful for age-dating), and that
Fomalhaut and TW PsA belong to the Castor group.  The question of
whether the CMG is actually useful for age-dating will await a future
investigation. For this study, I focus solely on the ages of Fomalhaut
and TW PsA, and assess the physicality of that binary.

\section{Analysis}

\subsection{Isochronal Age for Fomalhaut}

An isochronal age for Fomalhaut can be estimated through comparing its
\teff\, and luminosity to modern evolutionary tracks.  The stellar
parameters for Fomalhaut are fairly well determined due to its
brightness and proximity, which has enabled the star to have its
diameter measured interferometrically. Here I estimate refined HR
diagram parameters for Fomalhaut and estimate an isochronal age.

Basic stellar parameters for Fomalhaut are listed in Table
\ref{tab:data}. \citet{Davis05} estimated the bolometric flux of
Fomalhaut to be 8.96\,$\pm$\,0.25 $\times$10$^{-9}$ \wmsq, which I
adopt. Combining this with the revised Hipparcos parallax from
\citet{vanLeeuwen07} of $\varpi$ = 129.81\,$\pm$\,0.47 mas ($d$ =
7.704\,$\pm$\,0.028 pc), this results in a bolometric luminosity for
Fomalhaut of 16.63\,$\pm$\,0.48 L$_{\odot}$ or \logl\, =
1.221\,$\pm$\,0.013 dex\footnote{I adopt a revised solar luminosity of
  L$_{\odot}$ = 3.8270\,($\pm$\,0.0014) $\times$ 10$^{33}$ erg/s based
  on the total solar irradiance (TSI) of $S_{\odot}$ =
  1360.8\,($\pm$\,0.5) W\,m$^{-2}$ \citep{Kopp11} calibrated to the
  NIST radiant power scale, and the IAU 2009 value for the
  astronomical unit (149597870700\,$\pm$\,3 m). Using the bolometric
  magnitude zero-point proposed by IAU Commissions 25 and 36 of L =
  3.055 $\times$10$^{28}$ W, this translates to a solar absolute
  bolometric magnitude of M$_{bol}$ = 4.7554\,($\pm$0.0004) mag on
  that scale.  To force the recent TSI measurement to a scale where
  M$_{bol}$ = 4.75 \citep[a commonly adopted value;][]{Torres10}, the
  zero-point luminosity could be adjusted to 3.040$\times$10$^{28}$ W.
  One can calculate a modern \teff\, for the Sun by combining the new
  luminosity with the solar radius (695660 km) from \citet[][where I
  adopt $\pm$100 km error based on their discussion]{Haberreiter08}.
  The resultant solar \teff\, is 5771.8\,$\pm$\,0.7 K.}.
\citet{Absil09} resolved a small amount of K-band excess due to
circumstellar dust, and reported a revised limb-darkened diameter
taking into account all the available VLTI data: $\theta_{LD}$ =
2.223\,$\pm$\,0.022 mas.  Using the relation from \citet{vanBelle99}
(\teff\, = (2341 K) $\times$ $(f_{bol} / \theta^2_{LD})^{(1/4)}$,
where $f_{bol}$ is in units of 10$^{-8}$ erg cm$^{-2}$ s$^{-1}$ and
$\theta_{LD}$ is in mas) with the bolometric flux from \citet{Davis05}
and the limb-darkened diameter from \citet{Absil09}, I derive a new
\teff\, of 8590\,$\pm$\,73 K and radius 1.842\,$\pm$\,0.019
R$_{\odot}$.  The \teff\, is only slightly lower than recent estimates
\citep[e.g.][]{Davis05}.

\begin{deluxetable}{lccccc}
\tabletypesize{\scriptsize}
\setlength{\tabcolsep}{0.03in}
\tablewidth{0pt}
\tablecaption{Stellar Parameters\label{tab:data}}
\tablehead{
{(1)}           &{(2)}                 &{(3)}                &{(4)}   & {(5)}\\
{Value}         &{$\alpha$ PsA}        &{TW PsA}             &{Units} & {Ref.}
}
\startdata
$\alpha_{ICRS}$ & 344.411773           & 344.099277          & deg    & 1\\
$\delta_{ICRS}$ & -29.621837           & -31.565179          & deg    & 1\\
Parallax        &  129.81\,$\pm$\,0.47 & 131.42\,$\pm$\,0.62 & mas    & 1\\
Distance        &  7.704\,$\pm$\,0.028 & 7.609\,$\pm$\,0.036 & pc     & 1\\
\pmra\,         &  329.95\,$\pm$\,0.50 & 331.11\,$\pm$\,0.65 & \masyr & 1\\
\pmdec\,        & -164.67\,$\pm$\,0.35 &-158.98\,$\pm$\,0.48 & \masyr & 1\\
\vrad\,         &   6.5\,$\pm$\,0.5    & 6.6\,$\pm$\,0.1     & \kms   & 2,3\\
m$_V$           & 1.155\,$\pm$\,0.005  & 6.488\,$\pm$\,0.012 & mag    & 4\\
Period          & ...                  & 10.3                & day    & 5\\
Spec. Type      & A3 Va                & K4 Ve               & ...    & 6,7\\
\teff\,         & 8590\,$\pm$\,73 K    & 4594\,$\pm$\,80     & K      & 8,9\\
$f_{\rm bol}$   & 8.96\,$\pm$0.25      & 0.10075        & nW m$^{-2}$ & 10,9\\
U               & -5.71\,$\pm$\,0.16  & -5.69\,$\pm$\,0.06  & \kms  & 8\\
V               & -8.26\,$\pm$\,0.28  & -8.16\,$\pm$\,0.07  & \kms  & 8\\
W               &-11.04\,$\pm$\,0.38  &-10.96\,$\pm$\,0.08  & \kms  & 8\\
S$_{tot}$       & 14.92\,$\pm$\,0.33  & 14.80\,$\pm$\,0.07  & \kms  & 8\\
X$_{gal}$       &  3.06               &  3.14               & pc    & 8\\
Y$_{gal}$       &  1.14               &  0.90               & pc    & 8\\
Z$_{gal}$       & -6.98               & -6.88               & pc    & 8\\ 
Mass            & 1.92\,$\pm$\,0.02   & 0.73$^{+0.02}_{-0.01}$ & \msun & 8,9\\
M$_V$           & 1.722\,$\pm$\,0.009 & 7.081\,$\pm$\,0.016 & mag   & 8\\
Luminosity      & 16.63\,$\pm$\,0.48  & 0.189\,$\pm$\,0.013 & \lsun & 8\\
\logl           & 1.221\,$\pm$\,0.013 &-0.723\,$\pm$\,0.029 & dex   & 8\\
Radius          & 1.842\,$\pm$\,0.019 & 0.688\,$\pm$\,0.034 & \rsun & 8\\
\enddata
\tablecomments{References: 
(1) \citet{vanLeeuwen07},
(2) \citet{Gontcharov06}, 
(3) \citet{Nordstrom04}, 
(4) \citet{Mermilliod94}, 
(5) \citet{Busko78}, 
(6) \citet{GrayGarrison89A} (standard), 
(7) \citet{Keenan89}, 
(8) this paper (derived quantities discussed in \S3, using other
values in the table),
(9) \citet{Casagrande11},
(10) \citet{Davis05}. S$_{tot}$ is the barycentric speed. I assume \lsun\, =
  3.827 $\times$ 10$^{33}$ erg\,s$^{-1}$ (see footnote 1) and \rsun\,
  = 695660 km \citep{Haberreiter08}. 
}
\end{deluxetable}

\begin{figure}
\epsscale{1.0}
\plotone{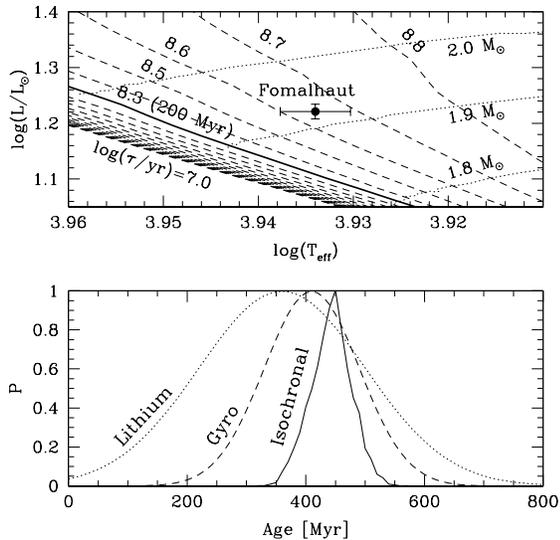}
\caption{{\it Top}: Theoretical HR diagram for Fomalhaut, with
  isochrones and evolutionary tracks from \citet{Bertelli08} (assuming
  protosolar composition of Y = 0.27, Z = 0.017).  Isochrones are in
  steps of 0.1 dex starting at log(age/yr) = 7.0.  A 200 Myr
  (log(age/yr) = 8.3) is plotted thick solid line.  {\it Bottom}:
  Normalized age probability distributions: {\it solid} is isochronal
  age for Fomalhaut using \citet{Bertelli08} tracks, {\it dashed} is
  gyrochronology age for TW PsA, and {\it dotted} is an approximate Li
  age through comparing TW PsA to open clusters (Sec. 3.3).
\label{hrd}}
\end{figure}

To calculate an isochronal age, I overlay the new HR diagram point for
Fomalhaut on the evolutionary tracks of \citet{Bertelli08} (Fig.
\ref{hrd}, top).  I assume that Fomalhaut has a chemical composition
similar to the proto-Sun, with an asteroseismically-motivated (and
diffusion corrected) composition of Y = 0.27 and Z = 0.017 \citep[see
][and references therein]{Serenelli10}\footnote{This is motivated by
  approximately solar metallicity of the companion TW PsA
  \citep{Barrado98,Casagrande11}.}.  I generate HR diagram positions
by Monte Carlo sampling the bolometric flux, parallax, and
limb-darkened radius values from their quoted values and uncertainties
(assuming a normal distribution), and interpolating within the
\citet{Bertelli08} tracks.  The \teff\, and luminosity of Fomalhaut
are consistent with a mass of 1.95\,$\pm$\,0.02 \msun\, and age of
433\,$\pm$\,36 Myr. The uncertainties only take into account
observational errors, and not systematic uncertainties in chemical
composition and input physics.  To estimate systematic uncertainties
due to different assumed solar composition and input physics, I
estimate calculate ages and masses for 3 more sets of tracks (see
Table \ref{tab:ages_iso}).  The expectation ages for the 4 sets of
tracks are very similar\footnote{Independently, the TYCHO evolutionary
  tracks yield an age of 476 Myr (Patrick Young, 2012, priv. comm.).},
and the mean of the ages from the 4 tracks is 450 Myr with a 22 Myr
(5\%) rms scatter (which is a reasonable estimate of the systematic
error considering slightly different assumed protosolar abundances and
input physics). Considering the typical observational error in age
($\pm$33 Myr; 7\%), this suggests a total isochronal age uncertainty
of $\pm$40 Myr (9\%).  For the four sets of tracks, the average mass
is 1.923 \msun, with $\pm$0.014 \msun\, rms (systematic error
component) and $\pm$0.016 \msun\, scatter due to the observational
errors. Note that this new mass (1.92\,$\pm$\,0.02 \msun) is similar
to previous estimates \citep[e.g.][]{Kalas08}, but 16\% lower than the
2.3 \msun\, quoted by \citet{Chiang09}.  The new estimated mass is in
line with observed trends in \teff\, and \logl\, vs. mass for main
sequence stars in eclipsing binaries \citep{Malkov07}, which
empirically predict $\sim$1.9-2.0 \msun.

\begin{deluxetable}{lllll}
\tabletypesize{\scriptsize}
\tablecaption{Age and Mass Estimates for Fomalhaut\label{tab:ages_iso}}
\tablewidth{0pt}
\tablehead{
\colhead{Tracks}   & {Y}   & {Z}    & \colhead{Age(Myr)} & \colhead{Mass}}
\startdata
\citet{Bertelli09} & 0.270 & 0.017  & 444$^{+32}_{-37}$ & 1.922\,$\pm$\,0.016\\
\citet{Marigo08}   & 0.273 & 0.019  & 427$^{+27}_{-32}$ & 1.911$^{+0.014}_{-0.016}$\\ 
\citet{Dotter08}   & 0.274 & 0.0189 & 478\tablenotemark{a} & 1.92\tablenotemark{a}\\   
\citet{Yi01}       & 0.264 & 0.017  & 453$^{+38}_{-32}$ & 1.943$^{+0.017}_{-0.014}$
\enddata
\tablenotetext{a}{I was unable to derive reliable uncertainties using 
these tracks.}
\end{deluxetable}

\subsection{Fomalhaut and TW PsA: A Physical Binary?}

Although overlooked in most recent literature on Fomalhaut, the star
has a likely stellar companion: TW PsA (GJ 879, HIP 113283).  That TW
PsA and Fomalhaut appear to share proper motion and parallax appears
to have been first noticed by \citet{Luyten38}. TW PsA is an active
K4Ve star \citep[][]{Keenan89}, at a projected separation of
1$^{\circ}$.96 ($\sim$7100''), and has been listed as among the widest
($\sim$50000 AU) candidate binaries known \citep{Gliese91}. The
proximity of TW PsA and its approximate co-motion with Fomalhaut led
to \citet{Barrado97} using TW PsA to age-date Fomalhaut.
\citet{Shaya11} included Fomalhaut and TW PsA as a wide binary in
their Bayesian search for multiple systems in the Hipparcos catalog,
and the system was one of only two binaries with separations $>$0.25
pc identified within 10 pc.

Given the utility of TW PsA to age-dating Fomalhaut, we should test
the physicality of the purported binary system using the best
available astrometry. The best available astrometric and radial
velocity data for Fomalhaut and TW PsA are listed in Table
\ref{tab:data}, and their degree of similarity is striking.  I adopt
the literature mean \vrad\, for Fomalhaut from \citet{Gontcharov06}
(6.5\,$\pm$\,0.5 \kms). This should reflect center-of-mass motion, as
the revised Hipparcos astrometric analysis was able to statistically
fit an unperturbed single-star solution for Fomalhaut's astrometry
\citep{vanLeeuwen07}. The astrometric acceleration from the original
Hipparcos reduction reported by \citet{Chiang09} is statistically
insignificant (1.7$\sigma$), and likely spurious.  For TW PsA, I adopt
the \vrad\, from \citet{Nordstrom04}, who reported a mean \vrad\, for
TW PsA of 6.6 \kms\, over 7 epochs over 3794 days. The star apparently
showed remarkable stability, with a quoted rms of $\pm$0.1 \kms.

I calculate the Galactic velocity vectors U, V, W (in the direction of
Galactic center, rotation, and North Galactic pole, respectively) from
the astrometry and radial velocities of Fomalhaut and TW PsA. The 3D
velocities are listed in Table \ref{tab:data}. The degree of
similarity is embarrassingly good, as their velocities agree within
$\pm$0.1 \kms\, in all three directions.  The barycentric speeds of
the star differ by only 0.1\,$\pm$0.5 \kms.  Just how remarkable is
the agreement in velocities?  I generated a catalog of 3D velocities
for 34,817 stars with updated Hipparcos astrometry from
\citet{vanLeeuwen07} (those with positive parallaxes) and literature
mean radial velocities from \citet{Gontcharov06}.  The only stars with
3D velocities within 1 \kms\, of Fomalhaut are TW PsA and HIP 3800
(0.99 \kms\, different). This suggests that $<$10$^{-4}$ of field
stars have velocities within 1\,\kms\, of Fomalhaut's, and that the
similarity in velocities between Fomalhaut and TW PsA (separated by
only $\sim$0.3 pc) is more than just a coincidence.

What is the 3D separation of Fomalhaut and TW PsA? Taking the parallax
values and uncertainties from \citet{vanLeeuwen07}, I generate
10$^{4}$ Monte Carlo realizations of the distances to Fomalhaut and TW
PsA (assuming $d$ = 1/parallax).  Fomalhaut has a parallax distance of
7.704\,$\pm$\,0.028 pc, while TW PsA is at 7.609\,$\pm$\,0.036 pc. The
3D separations in the simulations have a median separation of
0.280$^{+0.019}_{-0.012}$ pc (57.4$^{+3.9}_{-2.5}$ kAU; 68\%CL range
quoted). Many plausible binary systems are known with larger
separations \citep[e.g.][]{Shaya11}. Assuming that the census of the
nearest 100 star systems is
complete\footnote{http://www.recons.org/TOP100.posted.htm}, the local
density of star systems is 0.085 pc$^{-3}$. The chances of having an
unrelated star (system) within 0.28 pc of a nearby star is
approximately 1 in 130.

So the proximity in space and velocity between Fomalhaut and TW PsA
appear to be more than coincidental, but are they bound?  The escape
velocity from Fomalhaut (1.92\,\msun) at the separation of TW PsA is
0.21 \kms, suggesting that the observed difference in velocities
(0.1\,$\pm$\,0.5 \kms) is statistically consistent with the hypothesis
of TW PsA and Fomalhaut constituting a bound pair.  If one sets the
semi-major axis of TW PsA's orbit to be equal to the observed 3D
separation, and adopt a TW PsA mass of 0.73 \msun\,
\citep{Casagrande11}, then one estimates an orbital period of $\sim$8
Myr. The predicted amplitude of the orbital velocities would be 0.06
\kms\, for Fomalhaut and 0.15 \kms\, for TW PsA. 
 
\subsection{Age of TW PsA}

Rotation rates among late-F through early-M dwarfs appear to spin down
as they age through magnetically braking, approximately as rotation
period $\propto$ age$^{1/2}$ \citep{Skumanich72}.  Using the rotation
period from \citep[][10.3 days]{Busko78} and the gyrochronology curves
from \citet{Mamajek08}, and assuming $\pm$1.1 day rms fit to the gyro
relation, I estimate a gyrochronology age of 410\,$\pm$\,80 Myr.

Attempts to derive an isochronal age for TW PsA were discussed by
\citet{Barrado97}. Here I use the \teff\, and luminosity from Table 2
to derive new estimates of isochronal ages from evolutionary tracks.
Using the pre-MS evolutionary tracks of \citet{D'Antona97}, TW PsA is
consistent with having a mass of 0.71 \msun\, and an age of 52 Myr.
Using the \citet{Baraffe98} pre-MS tracks, TW PsA is consistent with
having a mass of 0.72 \msun\, and 66 Myr.  As discussed by
\citet{Barrado97} in their review of TW PsA's other youth diagnostics
(Li, activity), it is unlikely that the star is $<$100 Myr. The
spectroscopic surface gravity appears to be \logg\, $\simeq$ 4.5-4.7
\citep[e.g.][]{Dall05}, again consistent with a main sequence star.
In light of those findings, I consider the star to be a young main
sequence star, rather than pre-MS. Hence, I consider the pre-MS
isochronal age estimates to represent strict lower limits to TW PsA's
age (i.e. $>$50 Myr), rather than useful age estimates themselves.

TW PsA is also a coronal X-ray source, with L$_X$ = 10$^{28.33}$
\ergs, and fractional X-ray luminosity of \loglxlbol\, = -4.57
\citep{Wright11}. Although the star's color is slightly redder (B-V =
1.1) than the range probed by the calibrations in \citet{Mamajek08},
using the X-ray age relation from equation A3 of that paper, this
\loglxlbol\, value would be consistent with an age of $\sim$380 Myr.
Given the scatter in X-ray luminosities among stars in clusters of
similar mass \citep[$\sim$0.4 dex;][]{Mamajek08}, the age uncertainty
is approximately $^{+470}_{-220}$ Myr.

As pointed out by \citet{Barrado97}, TW PsA shows detectable Li (EW(Li
I $\lambda$6707) = 33$\pm$2 m\AA) consistent with an abundance of log
$N$(Li) = 0.6. The recent photometric Teff from \citet{Casagrande11}
that I adopt (\teff\, = 4594\,$\pm$\,80 K) is not far from the \teff\,
originally adopted by \citet{Barrado97} (4500 K). \citet{Barrado97}
plotted the Li I abundances for members of 4 clusters of different
ages (their Fig. 2; Pleiades, M34, UMa, Hyades), and given its
completeness, there is little reason to repeat the plot here. What
{\it has} changed in the past 15 years is the age scale for these
benchmark clusters. \citet{Barrado97} adopted the following age scale:
Pleiades - 85 Myr, M34 - 200 Myr, UMa - 300 Myr, Hyades - 700 Myr.
More recent evolutionary tracks are leading to slightly older ages
among the younger clusters. I adopt the following age scale: Pleiades
- 130 Myr \citep{Barrado04}, 220 Myr \citep{Meibom11}, UMa - 500 Myr
\citep{King03}, \& Hyades - 625 Myr \citep{Perryman98}. The Li
abundance for TW PsA appears to be intermediate between the Hyades and
Pleiades, and M34 and UMa. It is more Li-poor than the Pleiades and
M34 stars (hence $>$220 Myr), but more Li-rich than the UMa stars and
Hyades (hence $<$500 Myr). Based on the Li abundances alone, I adopt
an estimate of 360\,$\pm$\,140 Myr.

The three independent age estimates for TW PsA listed in Table
\ref{tab:ages} are consistent with a weighted mean age of
400\,$\pm$\,70 Myr.  This age estimate for TW PsA is independent of
any genetic association with Fomalhaut.

\begin{deluxetable}{ll}
\tabletypesize{\scriptsize}
\tablecaption{New Age Estimates for Fomalhaut \& TW PsA\label{tab:ages_new}}
\tablewidth{0pt}
\tablehead{
\colhead{Age (Myr)} & \colhead{Method}}
\startdata
450\,$\pm$\,40     & isochrones (Fomalhaut)\\
410\,$\pm$\,80     & rotation (TW PsA)\\
$>$50              & isochrones (TW PsA)\\
$\sim$380$^{+470}_{-220}$ & X-ray (TW PsA)\\
360\,$\pm$\,140    & Lithium (TW PsA)\\
{\bf 440\,$\pm$\,40} & {\bf final (both)}
\enddata
\end{deluxetable}

\section{Discussion}

The kinematic data are consistent with Fomalhaut and TW PsA comoving
within 0.1\,$\pm$\,0.5 \kms, and separated by only 0.28 pc. Given
their coincidence in position, velocity, and statistical agreement in
velocities expected for a wide bound binary, and remarkable agreement
among independent age estimates ($\sim$10\% agreement), I conclude
that Fomalhaut and TW PsA constitute a physical binary.  Therefore a
cross-comparison of their ages is useful.

The new age estimates for Fomalhaut and TW PsA are listed in Table
\ref{tab:ages_new}.  The new isochronal age for Fomalhaut
(450\,$\pm$\,40 Myr) is in good agreement with two recent isochronal
estimates: 480 Myr \citep{Rieke05} and 419\,$\pm$31 Myr
\citep{Zorec12}. It is clear that more modern evolutionary tracks and
constraints on the HR diagram position of Fomalhaut are leading to an
age twice as old as the classic age \citep[200 Myr;][]{Barrado97}.
Fig. \ref{hrd} (bottom) shows a pleasing overlap between the inferred
age probability distribution for Fomalhaut (using the
\citet{Bertelli08} tracks) and the gyrochronology and Li ages for TW
PsA (the two estimates with the smallest uncertainties).  Based on the
4 independent ages in Table \ref{tab:ages_new}, the rounded weighted
mean age for the Fomalhaut-TW PsA system is 440\,$\pm$\,40 Myr.  This
new estimate has relative uncertainties $\sim$5$\times$ smaller than
the age quoted by \citet{Barrado97} and \citet{Barrado98}
(200\,$\pm$\,100 Myr), and is tied to the contemporary open cluster
age scale and modern evolutionary tracks.

A factor of 2$\times$ older age for Fomalhaut has consequences for the
predicted brightnesses of substellar companions. Using the
\citet{Spiegel12} evolutionary tracks, it appears that a factor of
2$\times$ older age indicates that a given brightness limit at 4.5
$\mu$m (or M band) corresponds to thermal emission from a planet
roughly 2$\times$ as massive if it were 200 Myr. A 1 M$_{Jup}$ planet
of age 200 Myr has absolute magnitude $M_M$ = 20.4, but at 440 Myr is
approximately 1.2 magnitudes fainter ($M_M$ = 21.6). Future searches
for thermal emission from exoplanets orbiting Fomalhaut should take
into account this older age.

\acknowledgements

EEM acknowledges support from NSF award AST-1008908, and thanks Paul
Kalas, Mark Pecaut, Erin Scott, Tiffany Meshkat, and Matt Kenworthy
for comments on the manuscript, and the referee David Soderblom for a
helpful review.

%\bibliography{mamajek}{}

\bibliographystyle{apj}

\end{document}